\documentclass[10pt, aps, prd, preprintnumbers,nofootinbib]{revtex4-2}
\usepackage{graphicx}
\usepackage{dcolumn}
\usepackage{amsmath}
\usepackage{float}
\usepackage{makecell}
\usepackage{bm}
\usepackage{comment}
\usepackage{slashed}
\usepackage{gensymb}
\usepackage{tikz}
\usepackage[normalem]{ulem}
\usepackage[compat=1.1.0]{tikz-feynman}
\usepackage{ragged2e}
\usepackage[utf8]{inputenc}
\usepackage{tabularx}
\usepackage{array}
\usepackage{graphics}
\graphicspath{{}}
\usepackage{psfrag}
\usepackage{epsfig}
\usepackage{amssymb}
\usepackage{setspace}
\usepackage{rotating}
\usepackage{colortbl}
\usepackage{slashed}
\usepackage{enumitem}
\makeatletter
\usepackage{textcomp}
\usepackage[colorlinks=true,allcolors=purple]{hyperref}
\usepackage[capitalize]{cleveref}

\tikzset{
  graviton/.style={
    double,
    double distance=2pt,
    decorate,
    decoration={
      snake,
      amplitude=1.2pt,
      segment length=6pt
    }
  }
}

\newcommand{\stkout}[1]{\ifmmode\text{\sout{\ensuremath{#1}}}\else\sout{#1}\fi}

\begin{document}

\begin{flushright}
MI-HET-891
\end{flushright}

\title{
Intensity-Frontier Signals of Warped Extra Dimensions}

\author{Doojin Kim}
\email{doojin.kim@usd.edu}
\affiliation{
Department of Physics, University of South Dakota, Vermillion, SD 57069, USA
}
\author{Deepak Sathyan}
\email{dsathyan@tamu.edu}
\affiliation{Mitchell Institute for Fundamental Physics and Astronomy, Department of Physics  and Astronomy, Texas A$\&$M University, College Station, Texas 77843,  USA}

\author{Ankur Verma}
\email{averma1@tamu.edu}
\affiliation{
Department of Physics, University of South Dakota, Vermillion, SD 57069, USA
}

\date{\today}

\begin{abstract}
Can warped extra dimensions first appear at the intensity frontier rather than as TeV-scale resonances at colliders? We explore this possibility in extended warped models in which gravity propagates to a deep infrared region with warped scale $\Lambda_{\rm IR}\sim \mathcal{O}({\rm MeV})$, producing a densely spaced Kaluza-Klein (KK) graviton tower. We develop a benchmark photon-portal realization in which a visible vector sector reaches an intermediate GeV-scale brane, while the Higgs sector remains associated with a higher warped scale. The resulting graviton-photon couplings are controlled by wave-function overlap in the extra dimension, so the production rate is not governed simply by an independent mass and coupling as in conventional light-mediator simplified models. Instead, GeV-scale photons in beam-dump environments can preferentially produce heavier KK gravitons whose profiles probe the intermediate brane, after which the excited modes cascade down the tower. If decays into radion-like states are kinematically closed for the terminal mode, the lightest accessible KK graviton can be long-lived and decay visibly into a pair of photons. This leads to a distinctive intensity-frontier signature: heavy-mode production, intratower showering, and macroscopic electromagnetic decays. We present the model ingredients, derive the relevant overlap-controlled couplings, characterize the generic production and decay phenomenology, and discuss the theoretical and precision constraints on this class of low-scale warped scenarios.

\end{abstract}

\maketitle

\tableofcontents
\section{Introduction}\label{SectionI}

Warped extra dimensions were originally proposed as an elegant geometric solution to the electroweak hierarchy problem. In the Randall-Sundrum (RS) construction~\cite{Randall:1999ee,Randall:1999vf}, a five-dimensional anti-de Sitter spacetime is bounded by two branes, conventionally referred to as the ultraviolet (UV) and infrared (IR) branes. In the simplest realization, gravity propagates in the five-dimensional bulk, whereas the Standard Model (SM) fields, including the Higgs field, are localized near the IR brane. The exponential warp factor then redshifts mass parameters on the IR brane, allowing the weak scale to emerge from Planck-scale input parameters without introducing an exponentially small fundamental number.

The framework becomes even richer when the SM fields are allowed to propagate in the bulk~\cite{Davoudiasl:1999tf, Pomarol:1999ad, Grossman:1999ra, Chang:1999nh, Gherghetta:2000qt, Huber:2000ie}. In this case, the localization of fermion zero modes along the warped direction can generate hierarchical four-dimensional Yukawa couplings from order-one five-dimensional parameters, providing a geometric interpretation of the observed flavor hierarchies. This ``bulk SM'' version of the warped scenario, therefore, links the hierarchy problem to flavor physics (often called the flavor hierarchy problem) and has been one of the most attractive extensions of the original RS idea.

One of the most important phenomenological consequences of the bulk SM framework is the appearance of gauge and fermion Kaluza-Klein (KK) excitations with masses around the warped IR scale. When this scale is near the TeV range, energy-frontier colliders such as the Large Hadron Collider (LHC) provide the natural arena for testing the resulting new physics.
However, at the same time, precisely because those KK states couple to SM currents, these models are subject to strong constraints from precision electroweak observables, flavor and CP violation, and direct resonance searches. In minimal constructions, these constraints tend to push the KK scale well above the naive TeV scale, often beyond the most favorable discovery range of the LHC. Several mechanisms have been developed to relax these bounds, including enlarged custodial gauge symmetries that protect the electroweak $T$ parameter and the $Z b_L\bar b_L$ coupling~\cite{Agashe:2003zs, Carena:2006bn, Carena:2007ua}, as well as variations in the localization of fields~\cite{Hewett:2002fe, Carena:2004zn, Agashe:2004cp, Agashe:2006wa} and brane-localized kinetic terms (BLKTs)~\cite{Davoudiasl:2002ua, Carena:2002dz, delAguila:2003kd, Kobakhidze:2016jsr}.

A further development is the multi-brane warped framework~\cite{Agashe:2016rle}, in which different sectors of the theory can terminate at different positions in the warped extra dimension. In one realization, the matter and Higgs fields propagate only down to an intermediate brane, while gauge and gravity fields extend further into the infrared. This separation allows the sector most directly constrained by flavor and precision measurements to remain associated with a relatively high scale, while the gauge and gravitational sectors can access a deeper IR region. Such constructions have been studied primarily in the context of LHC signals from warped vector resonances and their cascade decays~\cite{Agashe:2016kfr, Agashe:2017wss, Agashe:2018leo, Agashe:2020wph}. In addition to LHC phenomenology, multi-brane warped constructions have been studied in a variety of other contexts, including the realization and stabilization of multiple hierarchical scales~\cite{Lee:2021wau}, radion dynamics and cosmological evolution~\cite{Cai:2021mrw,Girmohanta:2023sjv}, and dark-matter scenarios~\cite{Koutroulis:2024wjl}.

In this work, we take the same organizing principle in a different direction. Rather than asking only whether warped extra dimensions can produce TeV-scale resonances at energy-frontier colliders, we ask:
\begin{itemize}
    \item whether a multi-brane warped geometry can generate low-scale, weakly coupled new physics relevant for the intensity frontier, including neutrino, beam-dump, and fixed-target experiments;
    \item what ingredients are required for such constructions to remain consistent with existing electroweak and collider constraints; and
    \item what generic phenomenological features follow from the resulting low-scale KK towers and their couplings to the visible sector.
\end{itemize}
The basic idea is to separate the IR endpoints of the Higgs, visible gauge, and gravitational sectors. The Higgs sector remains tied to a TeV or multi-TeV scale, the visible gauge sector can extend to an intermediate scale, and the gravitational sector can propagate to a still deeper IR brane. This opens the possibility that the first accessible imprint of a warped extra dimension is not a conventional TeV resonance, but a light, densely spaced tower of states that couples to the visible sector through geometric overlap.

More concretely, for such extra-dimensional signals to arise in neutrino, beam-dump, or fixed-target environments, three ingredients are needed simultaneously: a portal that can be accessed with GeV-scale beam energies, a much lighter state that can propagate over macroscopic distances, and a visible decay channel through which the new sector can be reconstructed or constrained.
A deeper gravitational region in the warped bulk naturally supplies the second ingredient. If the gravitational sector extends to a brane with $\Lambda_{\rm IR}\sim{\rm MeV}$, the resulting KK graviton tower is densely spaced, and its lightest modes can be parametrically lighter than the typical energy transfer available at intensity-frontier facilities. By itself, however, such a light gravitational tower would be difficult to produce efficiently from SM states localized near the Higgs brane, because the relevant wave-function overlaps are highly suppressed. This motivates introducing an intermediate brane at a GeV-scale warped position, together with a bulk vector sector that reaches this brane and communicates with the light graviton tower. The Higgs sector can then remain tied to a higher warped scale, while the intermediate vector sector provides an accessible portal to the low-scale gravitational tower.

Fixed-target and neutrino facilities with GeV-scale particle beams, such as DarkQuest~\cite{Apyan:2022tsd}, DUNE~\cite{DUNE:2020lwj}, FASER~\cite{FASER:2018bac}, and SHiP~\cite{SHiP:2021nfo} provide a natural experimental arena for this class of extra-dimensional scenarios. The typical momentum transfer in such environments can lie in the MeV--GeV range, precisely where a low-scale KK graviton tower may contain many kinematically accessible modes. If the graviton tower couples appreciably to a bulk vector sector extending to an intermediate GeV brane, Gertsenshtein-like production (for example, via the SM photon)~\cite{Gertsenshtein:1962} can populate KK graviton states with masses comparable to the beam energy or momentum transfer.
The produced gravitons need not be the lightest states in the tower. More generically, GeV-scale production can access highly excited KK gravitons sitting above a dense set of lighter spin-2 states. These excited modes may then undergo intratower cascade decays, successively populating lighter KK gravitons. The terminal state of such a cascade depends on the spectrum of other light gravitational-sector degrees of freedom. In particular, if decays to radion pairs are kinematically open, they can dominate because of the IR-enhanced graviton-radion couplings. If, instead, the terminal graviton lies below the two-radion threshold, the radion channel is closed, and the lightest accessible spin-2 state can become long-lived. Its subsequent decay back to visible photons can then lead to displaced electromagnetic signatures. This combination of GeV-scale production, tower cascades, and long-lived electromagnetic decays is the basic phenomenological motivation for studying low-scale warped sectors at intensity-frontier experiments.

The central model-building question is how the visible sector communicates with the deep gravitational region. The most direct possibility is to allow a visible electroweak gauge field to propagate to an intermediate GeV brane. In the concrete realization studied in this paper, the hypercharge gauge field extends to the GeV brane, while the $SU(2)_L$ gauge sector remains tied to the Higgs brane. After electroweak symmetry breaking (EWSB), the physical photon inherits support in the intermediate region through its hypercharge component, thereby enhancing the overlap between visible photons and the low-scale KK graviton tower.

This construction, however, comes with an important consistency challenge. The same hypercharge field that provides the photon-graviton overlap also gives rise to a hypercharge KK tower. Without additional structure, these vector KK modes can couple too strongly to the Higgs and to SM fermions localized near the Higgs brane, leading to tension with precision electroweak observables and direct searches. We therefore introduce BLKTs for the hypercharge field. These terms distort the KK wave functions and can suppress their dangerous couplings near the Higgs brane, while preserving an appreciable overlap between the photon zero mode and the light gravitational tower.

There is also a second, more secluded realization in which the visible electroweak sector remains localized near the Higgs brane, while a dark $U(1)_D$ gauge field propagates to the intermediate brane and communicates with the SM through kinetic mixing. This dark-photon route may provide a useful alternative way to access the same low-scale gravitational tower while reducing direct electroweak constraints. In the present work, however, we focus on the minimal visible-sector realization based on bulk hypercharge and leave the dark $U(1)_D$ construction for future study.

The rest of this article is organized as follows:
In \cref{sec:ModelSetup} we introduce the extended warped geometry model realization that motivates our analysis. In \cref{sec:EWSB} we then study EWSB and identify the origin of the associated precision-electroweak tension. Next in \cref{sec:phenomodel}, we present the phenomenological model with BLKTs and show how it suppresses the dangerous hypercharge KK couplings while preserving an enhanced coupling of the KK graviton tower to visible photons. In \cref{sec:KKGSpectrum} and \cref{sec:couplings}, we then derive the KK graviton spectrum and the effective graviton-photon couplings relevant for phenomenology at intensity-frontier facilities. In \cref{sec:DUNEpheno} we eventually discuss the resulting photon-mediated production, intra-tower cascade-decays, and diphoton decay signatures of the lightest new particle. Finally, in \cref{sec:Conclusion}, we summarize our main results and conclusions.


\section{Model Setup}\label{sec:ModelSetup}
\begin{figure}[t]
\centering
\includegraphics[width=0.85\textwidth]{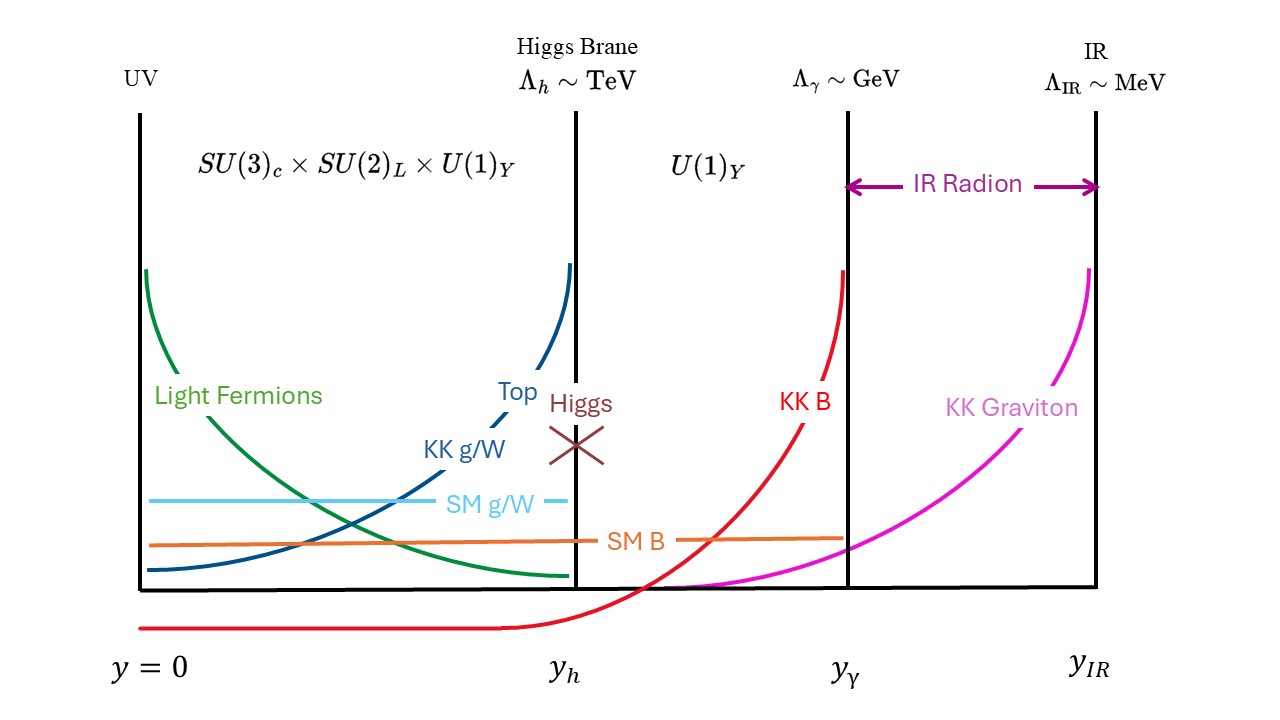}
\caption{Schematic illustration of the extended warped geometry considered in this work. The Higgs field is localized on the Higgs brane at $y_{\rm h}$, the $SU(2)_L$ gauge sector terminates there, the hypercharge gauge field propagates further down to the intermediate brane at $y_\gamma$, and gravity extends to the deepest infrared brane at $y_{\rm IR}$. 
}
\label{fig:setup}
\end{figure}

We consider an extended RS setup with multiple branes, described by the warped metric:
\begin{equation}
d s^2=e^{-2 \sigma(y)} \eta_{\mu \nu} d x^\mu d x^\nu+d y^2, \quad \sigma(y)=k|y|
\end{equation}
where the UV brane is located at $y=0$ while the deepest IR brane is located at $y=y_{\mathrm{IR}}$. In addition, we introduce two intermediate branes located at
\begin{equation}
0 < y_{\mathrm{h}} < y_{\gamma} < y_{\mathrm{IR}},
\end{equation}
and we schematically illustrate the overall extended warped geometry in Fig.~\ref{fig:setup}.
The intermediate branes allow different sectors of the theory to be localized at, or to probe, distinct warped scales within a single higher-dimensional geometry.\footnote{We work in the limit in which the intermediate branes define the support of the various sectors without appreciably modifying the background AdS$_5$ geometry. More precisely, the intermediate branes should be understood as effective sector-defining hypersurfaces in the phenomenological description used here. We assume that their gravitational backreaction is negligible compared with
that of the UV and deepest IR branes.} In particular, the brane positions define the warped scales
\begin{equation}
\Lambda_{\mathrm{h}} = k e^{-ky_{\mathrm{h}}},
\qquad
\Lambda_{\gamma} = k e^{-ky_{\gamma}},
\qquad
\Lambda_{\mathrm{IR}} = k e^{-ky_{\mathrm{IR}}}.
\end{equation}
Throughout this paper, we take the benchmark scale choices for them to be
\begin{equation}
\Lambda_{\mathrm{h}} \sim \mathcal{O}(\mathrm{TeV}), \qquad
\Lambda_{\gamma} \sim \mathcal{O}(\mathrm{GeV}), \qquad
\Lambda_{\mathrm{IR}} \sim \mathcal{O}(\mathrm{MeV}).
\end{equation}

The gravitational sector is allowed to propagate over the full interval $0 \le y \le y_{\mathrm{IR}}$, and therefore the KK graviton spectrum is controlled by the deepest IR scale $\Lambda_{\mathrm{IR}}$. In contrast, the Higgs doublet is localized on the usual TeV-scale 
brane or Higgs brane at $y=y_{\mathrm{h}}$,\footnote{Precision constraints generally favor a Higgs-brane scale of $\mathcal{O}(10~{\rm TeV})$ or higher. We use a TeV-scale Higgs brane in some benchmark estimates only for illustration; the qualitative conclusions do not rely on this choice.} ensuring that EWSB remains tied to $\Lambda_{\mathrm{h}}$ rather than the lower IR scales. The remaining model-building freedom lies in the vector sector, which determines how the SM communicates with the graviton tower through fields that extend into the intermediate GeV region.

As a concrete realization, we take only the hypercharge gauge field, associated with $U(1)_Y$, to propagate down to the intermediate GeV brane at $y=y_{\gamma}$, whereas the $SU(2)_L$ gauge fields are assumed to propagate only down to the Higgs brane at $y=y_{\rm h}$~\cite{Agashe:2017wss}. EWSB occurs on the Higgs brane, and the physical photon is therefore obtained, as usual, from the mixing of the neutral electroweak zero modes. The important consequence of this construction, following the logic of Ref.~\cite{Agashe:2017wss}, is that the bulk gauge field with support in the GeV region is the hypercharge field $B_\mu$. The photon then acquires support in the GeV region through its hypercharge component.

Another possible realization is to let the entire electroweak sector propagate only down to the Higgs brane, while allowing an additional dark or hidden gauge field, associated with $U(1)_D$, to propagate further down to an intermediate GeV brane at $y=y_D$. The graviton tower would then couple directly to the dark photon sector, with visible signatures induced through kinetic mixing with the ordinary photon. We leave this alternative construction and its phenomenological implications to forthcoming work.

Having specified the field content and the effective endpoints of the different sectors, we briefly comment on stabilization and on the scalar degrees of freedom associated with the extended geometry.
Stabilization of branes via Goldberger-Wise mechanism\footnote{Other stabilization mechanisms have been studied in RS-like models with MeV-scale sectors \cite{Csaki:2023pwy,Luo:2025alo}.}~\cite{Goldberger:1999uk} in a generic multi-brane scenario with $N$ 3-branes, in principle, requires $N-1$ independent radion modes corresponding to the independent inter-brane separations~\cite{Lee:2021wau,Donini:2025cpl,Girmohanta:2023sjv}. However, we work phenomenologically in the  weak-backreaction limit where the AdS curvature is taken to be approximately the same across the different bulk region, $k_i\simeq k$, so that the background can be approximately described by a single warp factor. In this limit, the intermediate-brane tension is negligible and the intermediate brane acts as weakly gravitating brane. Using the results from Ref.~\cite{Lee:2021wau} for the 3-brane setup, the radion associated mainly
with the intermediate branes has mass $m_{r_I}^2 \propto 1/(k_2-k_1)=1/{\delta k}$, and hence becomes parametrically heavy as
$\delta k\to0^+$.
We therefore integrate out this heavy mode and retain only
the lightest radion associated with the deepest IR interval.

\section{Electroweak Symmetry Breaking}
\label{sec:EWSB}
We first examine whether the model remains viable when only the hypercharge gauge field extends below the Higgs brane, while the $SU(2)_L$ gauge fields propagate only down to the Higgs brane at $y=y_{\rm h}$. EWSB provides the first nontrivial consistency test of this setup: the Higgs vacuum expectation value mixes the neutral $SU(2)_L$ zero mode with the hypercharge zero mode and the hypercharge KK tower, thereby generating brane-induced mass terms for the neutral gauge modes. The $SU(2)_L$ sector also contains KK excitations in principle, but their masses are controlled by the TeV scale $\Lambda_{\rm h}$ and hence they do not form part of the light neutral tower associated with the intermediate GeV brane. We therefore treat their effects as higher-scale corrections and focus on the mixing between the electroweak zero modes and the hypercharge KK tower, as well as on the resulting couplings of the neutral KK states to SM fields localized on, or toward the UV side of, the Higgs brane.

The purpose of this section is to identify the origin of the precision-electroweak tension and to motivate the BLKT construction introduced later.
Before EWSB, we use canonically normalized five-dimensional gauge fields, with the five-dimensional gauge couplings appearing in the covariant derivative. The relevant gauge action is
\begin{equation}
S_{\mathrm{gauge}} \supset
-\frac{1}{4 }\int d^4x\int_0^{y_h}dy \sqrt{-g}\, W_{MN}^a W^{aMN}
-\frac{1}{4}\int d^4x\int_0^{y_{\gamma}}dy \sqrt{-g}\, B_{MN} B^{MN}.
\end{equation}
The corresponding covariant derivative on the Higgs brane is given by
\begin{equation}
D_\mu = \partial_\mu
- \frac{i g_{5W}}{2} W_\mu^a(x,y_{\rm h})\sigma^a
- \frac{i g_{5Y}}{2} B_\mu(x,y_{\rm h}) ,
\end{equation}
where $\sigma^a$ denotes the usual Pauli matrices.
As discussed in the previous section, the different integration ranges reflect the central feature of the setup: the $SU(2)_L$ gauge fields propagate only up to the Higgs brane, whereas the hypercharge gauge field extends further into the bulk, down to the intermediate GeV-scale brane at $y=y_\gamma$.
For the flat zero modes, this gives the tree-level four-dimensional gauge couplings
\begin{equation}
g = \frac{g_{5W}}{\sqrt{y_h}},
\qquad
g' = \frac{g_{5Y}}{\sqrt{y_{\gamma}}}.
\end{equation}
The Higgs doublet is localized on the Higgs brane at $y=y_{\rm h}$.
Working in terms of the canonically normalized brane Higgs field, we write
\begin{equation}
S_H \supset \int d^4x
\left[
\left(D_\mu H\right)^\dagger \left(D^\mu H\right)
- \lambda \left(|H|^2 - \frac{v^2}{2}\right)^2
\right]_{y=y_{\rm h}},
\end{equation}
where 
\begin{equation}
\qquad
\langle H \rangle =
\frac{1}{\sqrt{2}}
\begin{pmatrix}
0 \\ v
\end{pmatrix}.
\end{equation}
The Higgs vacuum expectation value then generates the neutral brane mass term
\begin{equation}
\mathcal{L}_{\rm neutral}
=
\frac{v^2}{8}
\left[
g_{5W} W_\mu^3(x,y_{\rm h})
-
g_{5Y} B_\mu(x,y_{\rm h})
\right]^2 .
\end{equation}
This term mixes the neutral $SU(2)_L$ zero mode with the hypercharge zero mode and, because $B_\mu$ propagates beyond the Higgs brane, with the hypercharge KK tower.

At this stage, it is useful to distinguish the unmixed KK basis defined before
EWSB from the physical mass eigenbasis obtained
after diagonalization. We expand the neutral gauge fields in the pre-EWSB KK
basis as
\begin{equation}
W_\mu^3(x,y)
=
\frac{1}{\sqrt{y_{\rm h}}}
\sum_{n\geq 0} f_W^{(n)}(y)\, W_\mu^{3(n)}(x),
\qquad
B_\mu(x,y)
=
\frac{1}{\sqrt{y_\gamma}}
\sum_{n\geq 0} f_B^{(n)}(y)\, B_\mu^{(n)}(x),
\end{equation}
where the $W^3$ modes are defined on $0\leq y\leq y_{\rm h}$, whereas the
hypercharge modes are defined on $0\leq y\leq y_\gamma$. We take the mode
functions to be dimensionless, with flat zero-mode profiles normalized as
$f_W^{(0)}=f_B^{(0)}=1$. The Higgs-localized mass term then mixes all neutral
modes through their wave-function values at the Higgs brane.
It is convenient to define a column vector $\vec{V}$ and collect the corresponding Higgs-brane overlaps into
\begin{equation}
\vec V =
\left(
g f_W^{(0)}(y_{\rm h}), g f_W^{(1)}(y_{\rm h}), \cdots, -g' f_B^{(0)}(y_{\rm h}), -g' f_B^{(1)}(y_{\rm h}), \cdots
\right)^T.
\end{equation}
In the basis
\begin{equation}
\left(
W^{3(0)}, W^{3(1)}, \ldots,
B^{(0)}, B^{(1)}, \ldots
\right),
\end{equation}
the neutral gauge-boson mass matrix takes the compact form
\begin{equation}
M^2
=
\mathrm{diag}
\left(
0,(m_1^W)^2,\ldots,
0,(m_1^B)^2,\ldots
\right)
+
\frac{v^2}{4}\,\vec V\vec V^{\,T}.
\end{equation}
Equivalently, in components,
\begin{equation}
\begin{aligned}
\left(M^2\right)_{W_m W_n}
&=
(m_m^W)^2 \delta_{mn}
+
\frac{v^2}{4} g^2
f_W^{(m)}(y_{\rm h}) f_W^{(n)}(y_{\rm h}),
\\
\left(M^2\right)_{B_m B_n}
&=
(m_m^B)^2 \delta_{mn}
+
\frac{v^2}{4} g^{\prime 2}
f_B^{(m)}(y_{\rm h}) f_B^{(n)}(y_{\rm h}),
\\
\left(M^2\right)_{W_m B_n}
&=
-
\frac{v^2}{4} g g'
f_W^{(m)}(y_{\rm h}) f_B^{(n)}(y_{\rm h}) .
\end{aligned}
\end{equation}

After EWSB, the unbroken $U(1){\rm em}$ gauge symmetry guarantees the existence of a massless photon. Restricting first to the zero-mode sector, this photon corresponds to the neutral gauge-field combination orthogonal to the Higgs-induced mass term, giving the usual SM combinations
\begin{equation}
A_\mu^{(0)} = s_W W_\mu^{3(0)} + c_W B_\mu^{(0)},
\qquad
Z_\mu^{(0)} = c_W W_\mu^{3(0)} - s_W B_\mu^{(0)},
\end{equation}
where $s_W=\sin\theta_W$ and $c_W=\cos\theta_W$ with $\theta_W$ being the usual Weinberg angle. As in Ref.~\cite{Agashe:2017wss}, however, the gauge field with direct support at the intermediate GeV brane is the hypercharge field $B_\mu$, rather than a pure photon mass eigenstate. The photon therefore inherits overlap with the GeV region through its hypercharge component.
In principle, the Higgs-localized mass term also mixes the neutral KK towers. For the setup considered here, the $SU(2)_L$ KK modes live only on the shorter interval ending at $y=y_{\rm h}$ and have masses controlled by $\Lambda_{\rm h}\sim{\rm TeV}$. They are therefore much heavier than the hypercharge KK modes controlled by $\Lambda_\gamma\sim{\rm GeV}$, so their mixing with the light neutral tower is suppressed. We consequently treat the GeV-reaching neutral KK tower as the hypercharge KK tower to leading approximation. 

More explicitly, for the hypercharge KK profiles the $n$th mode can be written as
\begin{equation}
f_B^{(n)}(y)
=
\frac{e^{ky}}{N_n}
\left[
J_1\!\left(z_n^B(y)\right)
+\beta_n\,Y_1\!\left(z_n^B(y)\right)
\right],
\end{equation}
where $z_n^B(y)=(m_n^B/k)e^{ky}$, $m_n^B$ is the KK mass, $\beta_n$ is fixed by the boundary conditions, and $N_n$ is the normalization factor.
For the low-lying hypercharge KK modes, $z_n^B(y)\ll 1$ in the region near the UV and Higgs branes. Using the small-argument expansions of the Bessel functions,
\begin{equation}
J_1(z)\simeq \frac{z}{2},
\qquad
Y_1(z)\simeq -\frac{2}{\pi z}+\mathcal{O}(z\log z),
\end{equation}
one finds that the bulk solution becomes
\begin{equation}
f_B^{(n)}(y)
\simeq
\frac{e^{ky}}{N_n}
\left[
\frac{z_n^B(y)}{2}
-\frac{2\beta_n}{\pi z_n^B(y)}
+\cdots
\right]
=
\frac{1}{N_n}
\left[
-\frac{2\beta_n k}{\pi m_n^B}
+\mathcal{O}(e^{2ky})
\right].
\end{equation}
Thus near the UV and Higgs brane the hypercharge profile is a constant plus a subleading $e^{2ky}$ correction, not an exponentially suppressed function. This is precisely why precision constraints remain nontrivial even though only hypercharge reaches the GeV brane: the same feature that allows the hypercharge field to mediate between the visible sector and the GeV region also implies that its KK modes need not decouple from the Higgs brane.
For example, taking a representative GeV-scale mass, we find that the first hypercharge KK mode yields
\begin{equation}
z_1^B(y_h)
=
\frac{m_1^B}{k}e^{ky_h}
\sim
\frac{m_1^B}{\Lambda_h}
\sim
\frac{\mathrm{GeV}}{\mathrm{TeV}}
\sim 10^{-3},
\end{equation}
so the first mode is still deep in the small-$z$ regime at $y=y_h$. Its profile there is therefore approximately equal to the constant UV-side value,
\(
f_B^{(1)}(y_h)
\simeq
\text{constant},
\)
where the constant is fixed by the normalization and boundary conditions of the first KK eigenfunction. Thus the first-mode profile is not exponentially suppressed by a factor such as $e^{-k(y_{\gamma}-y_h)}\sim 10^{-3}$.

For a concrete benchmark with no BLKTs, we take $\Lambda_{\gamma}=1~{\rm GeV}$ and $\Lambda_{\rm h}=1~{\rm TeV}$. Solving the Neumann boundary conditions then gives
\begin{equation}
m_1^B \simeq 2.44~{\rm GeV},
\qquad
f_B^{(1)}(y_{\rm h})\simeq -0.178 .
\end{equation}
This result illustrates the point made above: near the higher-scale branes, the low-lying hypercharge KK profiles retain a non-negligible constant tail rather than being exponentially suppressed. The coupling of the first hypercharge KK mode to fields localized on the Higgs brane is therefore controlled directly by its wave-function value at $y=y_{\rm h}$. In the present normalization, and for negligible BLKTs, the hypercharge zero-mode profile is constant, $f_B^{(0)}(y_{\rm h})= 1$ , so that
\begin{equation}
\frac{g_1}{g_0}
\equiv
\frac{g_{B^{(1)}{\rm h}}}{g_{B^{(0)}{\rm h}}}
=
\frac{f_B^{(1)}(y_{\rm h})}{f_B^{(0)}(y_{\rm h})}
\simeq
f_B^{(1)}(y_{\rm h})
\simeq
-0.178 .
\end{equation}
The sign of this ratio is convention-dependent, while its magnitude shows that the first hypercharge KK mode couples to Higgs-brane matter at the level of roughly $18\%$ of the hypercharge zero-mode coupling. Thus, in the absence of additional model-building ingredients, this coupling is not suppressed by the warp hierarchy between the Higgs and GeV branes. This provides the motivation for introducing BLKTs, which can suppress the hypercharge KK profiles at the Higgs brane and thereby reduce the induced precision-electroweak tension.

\section{Phenomenological Model and Precision Constraints}
\label{sec:phenomodel}

As briefly discussed in the introduction, allowing the gravitational sector to propagate deeper into the warped bulk naturally gives rise to new states at lower energy scales. However, if the SM fields are localized near the Higgs brane, their wave-function overlap with the resulting light KK gravitons is extremely small, and the corresponding couplings are highly suppressed. In a multi-brane framework, this limitation can be alleviated by allowing a subset of SM fields, such as gauge fields, to propagate beyond the Higgs brane and thereby obtain a larger overlap with the KK graviton wave functions. A simple and phenomenologically viable realization is to let only the hypercharge gauge field propagate over the interval $0\leq y\leq y_\gamma$, while the remaining SM fields are localized on, or confined to the region above, the Higgs brane. After EWSB, the physical photon inherits this deeper support through its hypercharge component, enhancing the overlap between the light KK gravitons and the visible photon.

The price one pays is that the same hypercharge field also gives rise to a KK tower with non-negligible support near the Higgs brane, as discussed in the previous section. The resulting neutral KK vector states can mix with the electroweak zero modes and couple to fermions localized on, or residing above, the Higgs brane. Our model-building task is therefore twofold: we must retain the enhanced KK-graviton coupling to visible photons, which is essential for accelerator-based phenomenology, while suppressing the dangerous couplings of the hypercharge KK tower to the SM fields near the Higgs brane.

As summarized above, we now allow the bulk hypercharge gauge field to propagate down to the GeV brane at $y=y_{\gamma}$ and supplement its five-dimensional action with BLKTs at the endpoints of this interval. The relevant action is given by 
\begin{equation}
\label{eq:photon-blkt-action}
S_B=
-\frac{1}{4}\int d^4x\int_0^{y_{\gamma}}dy\,\sqrt{-g}\,
B_{MN}B^{MN}
-\frac{1}{4}\sum_{i\in\{0,y_{\gamma}\}}\int d^4x\,\sqrt{-g_{\mathrm{ind}}}\,
\frac{\delta_i}{k}\,B_{\mu\nu}B^{\mu\nu}\,
\end{equation}
where $\delta_0\equiv\delta_{\mathrm{UV}}$ and $\delta_{\gamma}$ are the BLKT coefficients on the UV and GeV branes, respectively.
Working in the gauge $B_5=0$ and $\partial_\mu B^\mu=0$, we expand
\begin{equation}
B_\mu(x,y)=\frac{1}{\sqrt{y_{\gamma}}}\sum_{n\ge 0}B_\mu^{(n)}(x)\,f_B^{(n)}(y). \label{eq:Bmuexp}
\end{equation}
The BLKTs modify the boundary conditions and normalization of the KK modes, but not the bulk equation of motion. The bulk profiles therefore satisfy
\begin{equation}
\left[-\partial_y^2+2k\,\partial_y\right]f_B^{(n)}(y)
=
e^{2ky}(m_n^B)^2 f_B^{(n)}(y),
\end{equation}
with the same Bessel-function form as in the absence of BLKTs,
\begin{equation}
f_B^{(n)}(y)
=
\frac{e^{ky}}{N_{B,n}}
\left[
J_1\!\left(z_n^B(y)\right)
+\beta_n^B\,Y_1\!\left(z_n^B(y)\right)
\right],
\qquad
z_n^B(y)=\frac{m_n^B}{k}e^{ky}.
\end{equation}
Here $N_{B,n}$ is fixed by the BLKT-modified normalization condition, while $\beta_n^B$ is determined by the endpoint boundary conditions.

Varying the action gives the BLKT-modified boundary conditions
\begin{align}
\left.\partial_y f_B^{(n)}\right|_{y=0}
&=
-\frac{\delta_{\rm UV}}{k}(m_n^B)^2 f_B^{(n)}(0),
\\
\left.\partial_y f_B^{(n)}\right|_{y=y_\gamma}
&=
\frac{\delta_\gamma}{k}(m_n^B)^2 e^{2k y_\gamma}
f_B^{(n)}(y_\gamma).
\end{align}
Equivalently, the UV and GeV-brane boundary conditions imply
\begin{align}
\beta_{\mathrm{UV}}^B(x_n^B)
&=
-\frac{\delta_{\mathrm{UV}}x_n^B J_1(x_n^B)+J_0(x_n^B)}
{\delta_{\mathrm{UV}}x_n^B Y_1(x_n^B)+Y_0(x_n^B)},
\\
\beta_{\gamma}^B(z_{\gamma,n}^B)
&=
-\frac{\delta_{\gamma}z_{\gamma,n}^B J_1(z_{\gamma,n}^B)-J_0(z_{\gamma,n}^B)}
{\delta_{\gamma}z_{\gamma,n}^B Y_1(z_{\gamma,n}^B)-Y_0(z_{\gamma,n}^B)},
\end{align}
where $x_n^B\equiv m_n^B/k$ and $z_{\gamma,n}^B\equiv z_n^B(y_{\gamma})=m_n^B/\Lambda_{\gamma}$.
The KK masses are obtained by requiring the two boundary conditions to determine the same coefficient,
\begin{equation}
\beta_{\mathrm{UV}}^B(x_n^B)=\beta_{\gamma}^B(z_{\gamma,n}^B).
\end{equation}

The BLKTs also modify the orthonormality condition, since they contribute
directly to the four-dimensional kinetic terms. In the convention above,
canonical normalization of the KK modes requires
\begin{equation}
\label{eq:blkt-norm}
\frac{1}{y_{\gamma}}
\left[
\int_0^{y_{\gamma}}dy\,
f_B^{(m)}(y)f_B^{(n)}(y)
+
\frac{\delta_{\rm UV}}{k}
f_B^{(m)}(0)f_B^{(n)}(0)
+
\frac{\delta_{\gamma}}{k}
f_B^{(m)}(y_{\gamma})f_B^{(n)}(y_{\gamma})
\right]
=
\delta_{mn}.
\end{equation}
For the zero mode, the bulk equation admits an exactly flat profile,
\begin{equation}
f_B^{(0)}(y)=N_{B,0}.
\end{equation}
The normalization condition then gives
\begin{equation}
N_{B,0}^{-2}
=
1+
\frac{\delta_{\rm UV}+\delta_{\gamma}}{k y_{\gamma}}.
\end{equation}

We are now ready to discuss precision constraints. The dominant precision-electroweak effects arise from two related sources: the mixing of the hypercharge KK tower with the neutral electroweak zero modes after EWSB, and the direct exchange of the resulting neutral KK vectors between SM currents. These effects modify the neutral-current sector, shifting observables such as $m_Z$, $\sin ^2 \theta_{\text {W}}$, $Z$-pole couplings, and asymmetries.

The mixing effect can be understood parametrically as follows. The $n$th hypercharge KK mode couples to the Higgs current with strength \(
g_n^H \sim g' f_B^{(n)}(y_h)
\)
up to normalization factors specified below. After the Higgs obtains a vacuum expectation value, this coupling induces a mass mixing between the SM-like zero-mode $Z$ boson\footnote{The physical $Z$ boson is identified not as the lightest massive neutral eigenstate, but as the eigenstate with the largest overlap with the usual zero-mode combination $Z_\mu^{(0)}=c_W W_\mu^{3(0)}-s_W B_\mu^{(0)}$.} and the hypercharge KK mode. Denoting this mixing by
\begin{equation}
\Delta_n^2 \sim \frac{v^2}{4}g_Z g_n^H = m_Z g_n^H v,
\qquad
g_Z=\sqrt{g^2+g'^2},
\end{equation}
the corresponding shift in the $Z$-boson mass is estimated as
\begin{equation}
\frac{\delta m_Z^2}{m_Z^2} \sim \sum_{n \geq 1} \frac{\left(g_n^H\right)^2 v^2}{{m_Z^2-}\left(m_n^B\right)^2}.
\end{equation}
In addition, virtual exchange of the neutral KK vectors generates effective four-fermion interactions. At momentum transfers below the KK-vector mass, integrating out the tower gives
\begin{equation}
\Delta \mathcal{L}_{\text {eff }} \sim \sum_{n \geq 1} \frac{\left(g_n^fg_n^{f'}\right)}{\left(m_n^B\right)^2}
\left(\bar{f} \gamma_\mu f\right)\left(\bar{f'} \gamma^\mu f'\right),
\end{equation}
where $g_n^{f}\propto g'\,f_B^{(n)}(y_f)$ denotes the couplings of the $n$th hypercharge KK mode to a fermion localized at $y=y_f$. 
Therefore, if the hypercharge KK tower begins at the GeV scale and the couplings $g_n^{H,f}$ are not sufficiently suppressed, both $Z$-pole observables and low-energy neutral-current measurements can receive unacceptably large tree-level corrections. The phenomenological requirement is therefore
\begin{equation}
g_n^{H},\,g_n^{f}\ll g'
\qquad\text{for the modes with}\qquad
m_n^B\sim \mathcal{O}(\mathrm{GeV}).
\end{equation}

With BLKTs, the hypercharge KK modes can remain light while their couplings to fields near the Higgs brane are parametrically reduced. The logic is similar to that of Ref.~\cite{Davoudiasl:2002ua}, although here it is applied to a setup in which the gauge interval terminates at an intermediate GeV brane rather than at the usual TeV brane.

For matter localized near $y=y_{\rm h}$, the coupling of the $n$th hypercharge KK mode is determined by the value of its wave function at the Higgs brane. With the KK expansion convention used above in Eq.~\eqref{eq:Bmuexp}, the coupling to Higgs-brane matter is
\begin{equation}
g_n^h
=
\frac{g_{5Y}}{{\sqrt{y_\gamma}}}\,
f_B^{(n)}(y_{\rm h}) .
\end{equation}
The observed hypercharge coupling is identified with the corresponding zero-mode coupling,
\begin{equation}
g_0\equiv g'
=
\frac{g_{5Y}}{\sqrt{y_\gamma}}\,
f_B^{(0)}(y_{\rm h}) .
\end{equation}
In the presence of BLKTs, the massless hypercharge zero mode remains flat in the bulk, but its normalization is modified. Defining
\begin{equation}
Z_0
\equiv
1+\frac{\delta_{\rm UV}+\delta_\gamma}{k y_\gamma},
\end{equation}
the normalized zero-mode profile is
\begin{equation}
f_B^{(0)}(y)=\frac{1}{\sqrt{Z_0}} .
\end{equation}
Thus the observed hypercharge coupling can be written as
\begin{equation}
g'
=
\frac{g_{5Y}}{\sqrt{y_\gamma Z_0}} .
\end{equation}
The coupling of an excited hypercharge KK mode to Higgs-brane matter, relative to the zero-mode coupling, is therefore
\begin{equation}
\frac{g_n^h}{g'}
=
\frac{f_B^{(n)}(y_{\rm h})}{f_B^{(0)}(y_{\rm h})}
=
\sqrt{Z_0}\, f_B^{(n)}(y_{\rm h}) .
\end{equation}
This expression makes clear how BLKTs can suppress the dangerous couplings. First, the BLKTs can distort the excited-mode wave functions so that $f_B^{(n)}(y_{\rm h})$ is small. Second, if moderately negative BLKTs are allowed, the zero-mode normalization factor $Z_0$ can be made positive but parametrically small through a partial cancellation between the bulk and brane contributions. Keeping the observed coupling $g'$ fixed then requires
\begin{equation}
g_{5Y}=g'\sqrt{y_\gamma Z_0},
\end{equation}
so the excited-mode couplings inherit an additional suppression proportional to $\sqrt{Z_0}$, up to the mode-dependent wave-function factor.

This negative-BLKT regime must be treated with care. One must remain in the region $Z_0>0$ so that the photon zero mode has a positive kinetic term, and one must avoid choices of BLKT coefficients that generate spurious light states or ghostlike modes in the KK spectrum. Within the viable region, however, the KK masses remain controlled by the GeV brane scale $\Lambda_\gamma$, while the dangerous couplings of the hypercharge KK tower to Higgs-brane fields can be suppressed both by wave-function distortion and by the additional $\sqrt{Z_0}$ factor. This is what makes it plausible to lower the endpoint of visible gauge propagation to the GeV brane without immediately violating precision-electroweak constraints. At the same time, the hypercharge tower can still couple to the KK graviton sector through the overlap integrals discussed in \cref{sec:couplings}.

\section{KK Graviton Spectrum}
\label{sec:KKGSpectrum}
The gravitational sector propagates over the full warped interval, down to the deepest brane at $y=y_{\mathrm{IR}}$. We start from the five-dimensional Einstein-Hilbert action
\begin{equation}
S_G=\frac{M_5^3}{4}\int d^4x\,dy\,\sqrt{|G|}\,R^{(5)}.
\end{equation}
The spin-2 KK spectrum is obtained by expanding the metric around the warped background as
\begin{equation}
G_{\mu\nu}(x,y)
=
e^{-2ky}
\left[
\eta_{\mu\nu}
+
2M_5^{-3/2}h_{\mu\nu}(x,y)
\right],
\end{equation}
where $h_{\mu\nu}$ denotes the transverse-traceless tensor fluctuation. We decompose it into four-dimensional KK modes as
\begin{equation}
h_{\mu\nu}(x,y)
=
\sum_{n=0}^{\infty}
h_{\mu\nu}^{(n)}(x)\,
f_G^{(n)}(y),
\end{equation}
where $h_{\mu\nu}^{(n)}(x)$ is the $n$th four-dimensional spin-2 mode and $f_G^{(n)}(y)$ is its wave function in the extra dimension.

For the massive modes, the profile equation is given by
\begin{equation}
\partial_y
\left(
e^{-4ky}\partial_y f_G^{(n)}
\right)
+
e^{-2ky}
\left(m_n^G\right)^2
f_G^{(n)}
=
0 .
\end{equation}
It is convenient to factor out the warp-factor dependence. Introducing the following variable
\begin{equation}
z_n^G(y)
=
\frac{m_n^G}{k}e^{ky},
\end{equation}
the equation reduces to a Bessel equation of order two. The general solution is
\begin{equation}
f_G^{(n)}(y)
=
\frac{e^{2ky}}{N_{G,n}}
\left[
J_2\!\left(z_n^G(y)\right)
+
\alpha_n^G
Y_2\!\left(z_n^G(y)\right)
\right],
\end{equation}
where $N_{G,n}$ is a normalization constant and $\alpha_n^G$ is fixed by the boundary conditions.

The allowed masses $m_n^G$ are determined by the boundary conditions at the UV and IR branes. 
In the absence of brane-localized curvature terms, the transverse-traceless tensor modes satisfy Neumann boundary conditions at the UV and IR branes,
\begin{equation}
\left.
\partial_y f_G^{(n)}
\right|_{y=0}
=
\left.
\partial_y f_G^{(n)}
\right|_{y=y_{\rm IR}}
=
0 .
\end{equation}
Using the Bessel-function solution, these boundary conditions imply
\begin{equation}
\alpha_{\rm UV}^G(m_n^G)
=
-
\frac{
J_1\!\left(z_n^G(0)\right)
}{
Y_1\!\left(z_n^G(0)\right)
},
\qquad
\alpha_{\rm IR}^G(m_n^G)
=
-
\frac{
J_1\!\left(z_n^G(y_{\rm IR})\right)
}{
Y_1\!\left(z_n^G(y_{\rm IR})\right)
}.
\end{equation}
The KK masses are determined by requiring that the two boundary conditions fix the same coefficient,
\begin{equation}
\alpha_{\rm UV}^G(m_n^G)
=
\alpha_{\rm IR}^G(m_n^G).
\end{equation}

For a large warp hierarchy, $z_n^G(0)=m_n^G/k\ll1$, and the mass eigenvalues are well approximated by the zeros of $J_1$ at the IR brane:
\begin{equation}
z_n^G(y_{\rm IR})
=
\frac{m_n^G}{k}e^{k y_{\rm IR}}
\simeq
x_n^G,
\qquad
J_1(x_n^G)=0 .
\end{equation}
Thus
\begin{equation}
m_n^G
\simeq
x_n^G\,k e^{-k y_{\rm IR}}
\equiv
x_n^G\,\Lambda_{\rm IR},
\end{equation}
where
\(
\Lambda_{\rm IR}
\equiv
k e^{-k y_{\rm IR}} .
\)
Numerically, $x_1^G\simeq3.83$, $x_2^G\simeq7.02$, and for large $n$,
\begin{equation}
x_n^G
\simeq
\left(n+\frac14\right)\pi .
\end{equation}
The KK graviton mass spacing is therefore set by the deepest IR scale,
\begin{equation}
\Delta m_G
\sim
\pi \Lambda_{\rm IR}.
\end{equation}
For $\Lambda_{\rm IR}$ in the MeV range, the KK graviton tower is correspondingly spaced by a few MeV.

The profile normalization is fixed by requiring the four-dimensional spin-2 fields to have canonically normalized kinetic terms. With the interval convention used here, this condition may be written as
\begin{equation}
\int_0^{y_{\rm IR}}dy\,
e^{-2ky}
f_G^{(m)}(y)f_G^{(n)}(y)
=
\delta_{mn},
\end{equation}
up to an overall convention-dependent factor that can be absorbed into the definition of $f_G^{(n)}$ and the corresponding coupling scale. The normalized profiles enter the overlap integrals that determine the couplings of KK gravitons to bulk gauge fields and to matter localized at different positions in the warped dimension.

\section{Effective KK Graviton Couplings to Bulk Vectors}
\label{sec:couplings}

We now derive the effective four-dimensional couplings between the light KK graviton tower and the visible photon sector. In the present multi-brane setup, the bulk vector field that extends down to the intermediate brane at $y=y_\gamma$ is the hypercharge gauge field. Therefore, the enhanced coupling of light KK gravitons to visible photons is inherited primarily from their overlap with the hypercharge component of the photon. This provides a generic mechanism by which a low-scale gravitational sector can communicate with visible electromagnetic degrees of freedom, without committing to a specific experimental realization.

The interaction follows from expanding the bulk hypercharge action, including
the hypercharge BLKTs, to linear order in the metric perturbation. Following
the logic of Ref.~\cite{Kobakhidze:2016jsr}, but adapting the result to a
hypercharge sector that terminates at $y=y_\gamma$ rather than at the deepest
IR brane, we write
\begin{equation}
S_{\rm int}
\supset
-\frac{1}{M_5^{3/2}}
\int d^4x \int_0^{y_\gamma}dy\,
h_{\mu\nu}(x,y)\,
T_B^{\mu\nu}(x,y).
\end{equation}
Here $T_B^{\mu\nu}$ denotes the stress tensor of the bulk hypercharge field,
including the endpoint contributions associated with the BLKTs.

Using the zero-mode expansion
\begin{equation}
B_\mu(x,y)
=
\frac{1}{\sqrt{y_\gamma}}\,
B_\mu^{(0)}(x)\,
f_B^{(0)}(y)
+\cdots,
\end{equation}
and the fact that the hypercharge zero mode is flat in the bulk,
\begin{equation}
f_B^{(0)}(y)=N_{B,0},
\qquad
N_{B,0}^{-2}
=
Z_0
=
1+\frac{\delta_{\rm UV}+\delta_\gamma}{k y_\gamma},
\end{equation}
one obtains an effective coupling between the $n$th KK graviton and two
hypercharge zero modes. Projecting onto the physical photon through
\(
B_\mu^{(0)}
=
c_W A_\mu^{(0)} - s_W Z_\mu^{(0)},
\)
the hypercharge contribution to the graviton-photon-photon coupling is
\begin{equation}
{\cal L}_{\rm int}
\supset
-\sum_n
\frac{1}{\Lambda_{\gamma\gamma,B}^{(n)}}\,
h_{\mu\nu}^{(n)}\,
T_\gamma^{(0)\mu\nu},
\end{equation}
with
\begin{equation}
\label{eq:cgn-general}
\frac{1}{\Lambda_{\gamma\gamma,B}^{(n)}}
=
\frac{c_W^2 N_{B,0}^2}{M_5^{3/2}y_\gamma}
\left[
\int_0^{y_\gamma}dy\, f_G^{(n)}(y)
+
\frac{\delta_{\rm UV}}{k}f_G^{(n)}(0)
+
\frac{\delta_\gamma}{k}f_G^{(n)}(y_\gamma)
\right].
\end{equation}
Equivalently, using the zero-mode normalization,
\begin{equation}
\frac{1}{\Lambda_{\gamma\gamma,B}^{(n)}}
=
\frac{c_W^2}{M_5^{3/2}
\left[
y_\gamma+\frac{1}{k}(\delta_{\rm UV}+\delta_\gamma)
\right]}
\left[
\int_0^{y_\gamma}dy\, f_G^{(n)}(y)
+
\frac{\delta_{\rm UV}}{k}f_G^{(n)}(0)
+
\frac{\delta_\gamma}{k}f_G^{(n)}(y_\gamma)
\right].
\end{equation}
The photon stress tensor is
\begin{equation}
T_{\gamma,\alpha\beta}^{(0)}
=
\frac{1}{4}\eta_{\alpha\beta}
F_{\mu\nu}^{(0)}F^{(0)\mu\nu}
-
F_{\alpha\nu}^{(0)}
{F^{(0)}}_{\beta}{}^\nu.
\end{equation}
Here Eq.~\eqref{eq:cgn-general} is the direct analogue of the overlap formula in Ref.~\cite{Kobakhidze:2016jsr}, generalized to a truncated hypercharge interval and to hypercharge BLKTs. The important feature is that the graviton profile is integrated only over the region explored by the corresponding gauge field. Thus, by moving the endpoint of the hypercharge interval from the Higgs brane down to the intermediate brane, the photon zero mode inherits sensitivity to a more infrared portion of the KK graviton wave functions through its hypercharge component.

Strictly speaking, the physical photon also contains an $SU(2)_L$ component,
\(
W_\mu^{3(0)}
=
s_W A_\mu^{(0)} + c_W Z_\mu^{(0)} .
\)
Since the $SU(2)_L$ gauge field propagates only down to the Higgs brane at $y=y_h$, this component gives an additional contribution
\begin{equation}
\frac{1}{\Lambda_{\gamma\gamma,W}^{(n)}}
=
\frac{s_W^2}{M_5^{3/2}y_h}
\int_0^{y_h}dy\, f_G^{(n)}(y),
\end{equation}
up to possible $SU(2)_L$ BLKT terms if present. The full photon coupling is
therefore
\begin{equation}
\frac{1}{\Lambda_{\gamma\gamma,{\rm eff}}^{(n)}}
=
\frac{1}{\Lambda_{\gamma\gamma,B}^{(n)}}
+
\frac{1}{\Lambda_{\gamma\gamma,W}^{(n)}} .
\end{equation}
In the parameter regime of interest, however, the novel enhancement comes from the hypercharge contribution because only the hypercharge field probes the intermediate GeV brane. The $SU(2)_L$ contribution is controlled by the shorter interval ending at $y_h$ and is therefore analogous to the usual TeV-brane-localized contribution.

The qualitative behavior of the overlap can be understood from the argument of the graviton profile at the hypercharge brane,
\begin{equation}
z_n^G(y_\gamma)
=
\frac{m_n^G}{k}e^{k y_\gamma}
=
\frac{m_n^G}{\Lambda_\gamma}.
\end{equation}
For modes with
\begin{equation}
m_n^G \ll \Lambda_\gamma,
\end{equation}
one has $z_n^G(y_\gamma)\ll1$, and the graviton profile remains on its small-argument tail throughout the hypercharge interval. The overlap with the photon zero mode is then suppressed. By contrast, for modes with
\begin{equation}
m_n^G \sim \Lambda_\gamma,
\end{equation}
the argument satisfies $z_n^G(y_\gamma)\sim{\cal O}(1)$, and the graviton profile develops appreciable support near the endpoint of the hypercharge interval. This is the regime in which the photon-graviton overlap is maximized.

This behavior is a generic consequence of the multi-brane construction. A bulk vector that terminates at an intermediate scale $\Lambda_\gamma$ couples most efficiently to the part of the KK graviton tower with masses near that scale. Lowering the endpoint of visible gauge propagation from the Higgs brane to an intermediate GeV brane therefore shifts the most strongly coupled portion of the graviton tower from the TeV scale down to the GeV scale. This provides a model-building route to low-scale new physics associated with a densely spaced KK graviton tower, while keeping the Higgs sector localized at a higher warped scale.

\begin{figure}[t]
\centering 
\includegraphics[width=.48\textwidth]{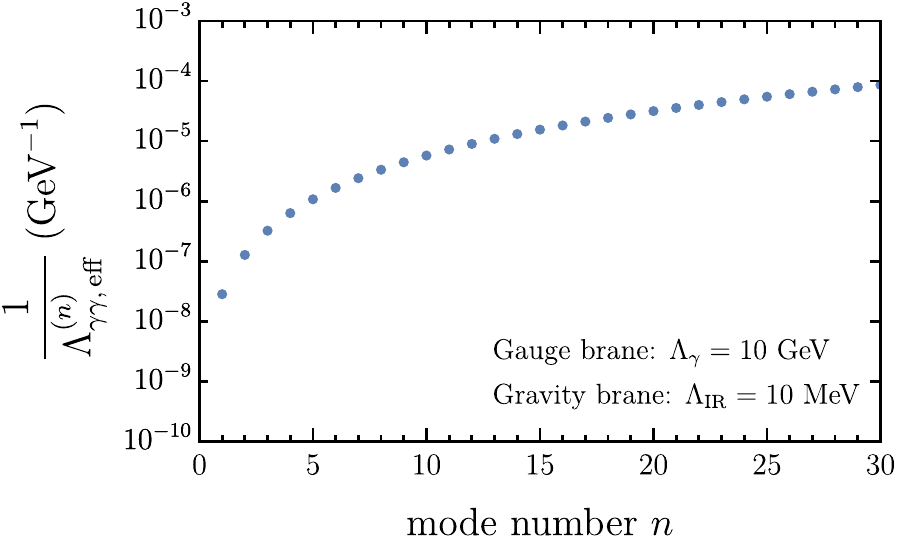}
\includegraphics[width=.495\textwidth]{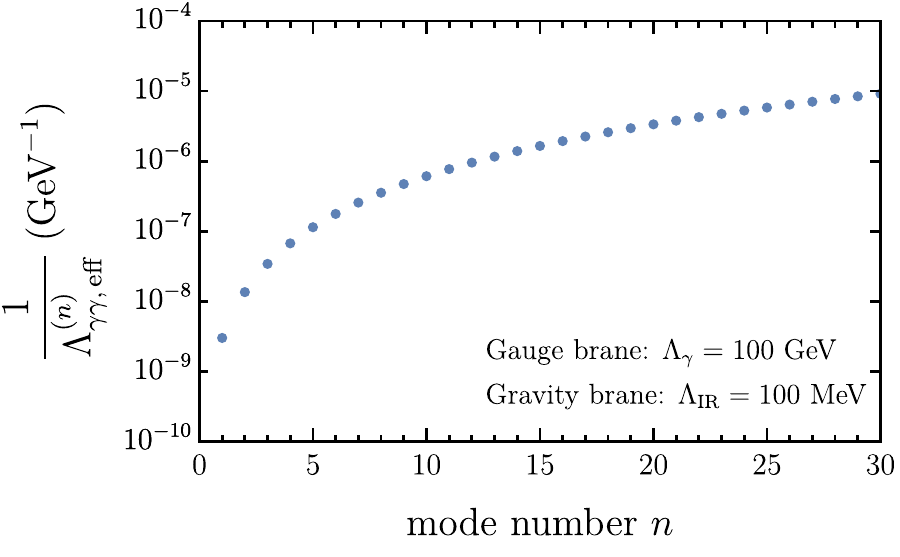}
\hfill

\caption{\label{fig:coupling} 
KK graviton couplings to photons from the normalized overlap of the KK graviton wave function with the photon zero-mode kinetic profile. A negative UV BLKT, $\delta_{\rm UV}=(-1+10^{-4})k y_\gamma$, is included while the total photon kinetic normalization remains positive.
}
\end{figure}

In Fig.~\ref{fig:coupling}, we illustrate the effective photon coupling as a function of the KK graviton mode number.\footnote{The numerical choice used in Fig.~\ref{fig:coupling} is intended as an illustrative benchmark showing how the overlap pattern changes when the photon kinetic profile is weighted toward the intermediate region. It should not be interpreted as a unique or optimized parameter point. More generally, the effective coupling depends on both the BLKT-induced zero-mode normalization and the detailed shape of the KK graviton wave function, while consistency requires a positive total kinetic normalization and the absence of ghostlike or spurious vector
modes.} At small $n$, the corresponding graviton masses lie well below the scale $\Lambda_\gamma$, so the overlap with the hypercharge interval is weak and the coupling is suppressed. As the mode number increases and $m_n^G$ approaches $\Lambda_\gamma$, the coupling grows because the graviton profile develops appreciable support near the hypercharge brane. At still larger $n$, the profile becomes increasingly oscillatory over the finite interval probed by the hypercharge field, so the weighted overlap need not increase monotonically.

\section{Phenomenology}\label{sec:DUNEpheno}

We now turn to the generic phenomenology of the low-scale KK graviton tower at intensity-frontier facilities. The key questions are how the new states can be produced in high-intensity environments and how their experimental signatures can be detected. In the framework developed above, the spin-2 KK modes can be light enough to be kinematically accessible in beam-based experiments, while their leading visible coupling arises through the SM photon. We therefore focus in this section on photon-mediated production and decay of KK gravitons, leaving detailed studies of specific experimental configurations to future work. 

\subsection{Production}

\begin{figure}
    \centering
\begin{tikzpicture}
\begin{feynman}

\vertex (gamin) at (-3,1.2) {$\gamma$};
\vertex[dot] (v1) at (0,1.2) {};
\vertex (grout) at (3,1.2) {$G_n^{\rm KK}$};

\vertex (Nin) at (-3,-1.2) {$N$};
\vertex[dot] (v2) at (0,-1.2) {};
\vertex (Nout) at (3,-1.2) {$N$};

\diagram*{
  (gamin) -- [photon] (v1),
  (v1) -- [graviton] (grout),
  (v1) -- [photon, edge label'={$\gamma^\ast$}] (v2),
  (Nin) -- [fermion] (v2) -- [fermion] (Nout),
};

\end{feynman}

\end{tikzpicture}
    \caption{Production of KK gravitons through a Gertsenshtein-like conversion of photons produced in the beam target. The conversion occurs in the electromagnetic field of the target nuclei at a fixed-target or beam-dump experiment.}
    \label{fig:prodKK}
\end{figure}
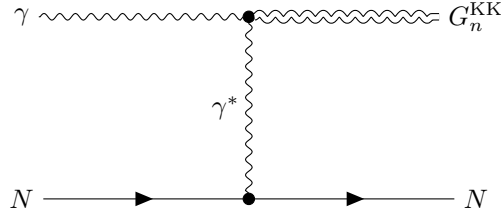

In a typical fixed-target experiment, an intense primary beam impinges on a target or beam dump and produces a large flux of secondary particles. Among these secondaries, photons are produced copiously, primarily through the electromagnetic shower initiated by charged particles in the target material. The photon flux can then source KK gravitons through a Gertsenshtein-like conversion process in the electromagnetic field of the target nuclei,
\begin{equation}
\gamma + T \to G_n + T,
\end{equation}
where $T$ denotes a target nucleus of mass $m_T$, as shown in Fig.~\ref{fig:prodKK}.
Since typical momentum transfers through the $t$-channel off-shell photon are small, the target nucleus can absorb the recoil coherently. The production amplitude, therefore, receives the usual coherent nuclear enhancement proportional to the nuclear charge $Z$, leading to a cross-section enhancement proportional to $Z^2$ as discussed shortly. 
The production rate is controlled by the effective graviton-photon coupling derived in \cref{sec:couplings}. Since the secondary photons in GeV-scale beam-dump or fixed-target environments typically carry energies of order a few GeV, the production process is most sensitive to KK gravitons with masses in the same range. This is also the mass range in which the overlap between the photon zero mode and the KK graviton wave function is enhanced in the present construction. Consequently, production is expected to favor GeV-scale KK gravitons over the much lighter MeV-spaced modes, whose profiles remain on the small-argument tail over the photon interval and hence couple more weakly to photons.

For a fixed incoming photon energy $E_\gamma$, the mode-by-mode production cross section can be written as
\begin{equation}
\sigma_n(E_\gamma)
=
\int_{t_-}^{t_+} dt\,
\frac{d\sigma(\gamma T\to G_n T)}{dt},
\end{equation}
where the two kinematic endpoints $t_\pm$ are given by
\begin{equation}
t_\pm
=
\left(m_n^G\right)^2
-
\frac{s-m_T^2}{2s}
\left[
s+\left(m_n^G\right)^2-m_T^2
\mp
\lambda^{1/2}
\left(
s,\left(m_n^G\right)^2,m_T^2
\right)
\right],
\end{equation}
with $s=m_T^2+2m_T E_\gamma$ and $\lambda(a,b,c)=a^2+b^2+c^2-2ab-2ac-2bc$.
Here \(t_+\) corresponds to the forward-scattering endpoint, where the
momentum transfer is smallest in magnitude.
In the forward, heavy-target regime, $|t|\ll \left(m_n^G\right)^2 \ll E_\gamma m_T$, the differential cross section takes the approximate form
\begin{equation}
\frac{d \sigma}{d t} \simeq \frac{e^2 Z^2}{32} \frac{\left(m_n^G\right)^4}{E_\gamma^2\left(\Lambda_{\gamma \gamma, \mathrm{eff}}^{(n)}\right)^2} \frac{F_T^2(t)}{t^2},
\end{equation}
where $F_T(t)$ is the nuclear form factor, which may be modeled, for example, by the Helm parameterization~\cite{Helm:1956zz}.
The essential model dependence enters through the effective graviton-photon coupling $1/\Lambda_{\gamma\gamma,{\rm eff}}^{(n)}$ and through the KK graviton mass $m_n^G$. Modes whose wave functions have larger overlap with the hypercharge, and hence photon, interval are produced more efficiently.

\begin{figure}[t]
\centering 
\includegraphics[width=.48\textwidth]{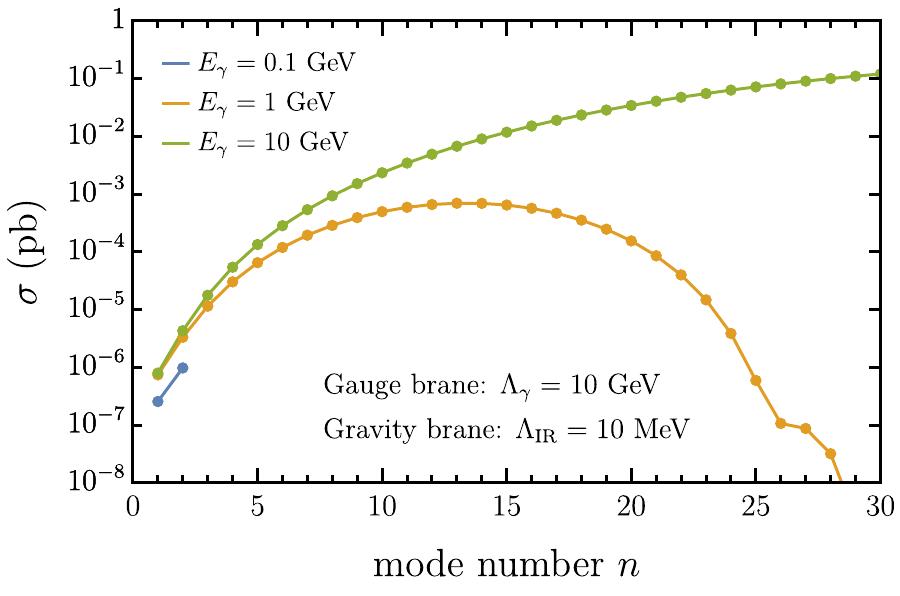}
\includegraphics[width=.48\textwidth]{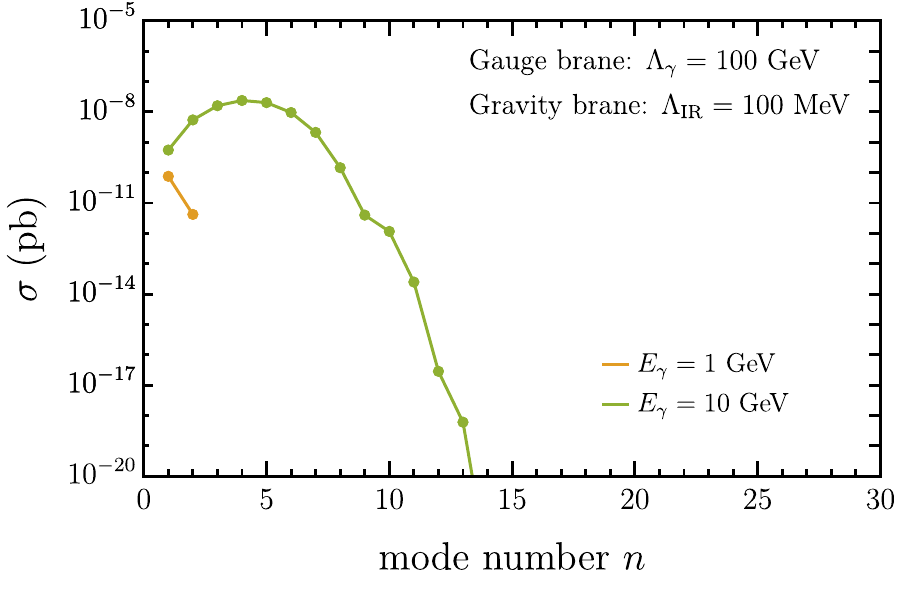}
\hfill

\caption{\label{fig:prod_xsec_mode} 
Mode-by-mode Gertsenshtein-like production cross section for the $n$th KK graviton, $\gamma N \rightarrow G_n N$, as a function of the KK mode number $n$, for incident photon energies $E_\gamma=0.1,1$, and 10 GeV. The left (right) panel corresponds to $\Lambda_\gamma=10(100) \mathrm{GeV}$ and $\Lambda_{\mathrm{IR}}=10(100) \mathrm{MeV}$. Only kinematically accessible modes are shown. The mode dependence reflects the interplay among the graviton-photon overlap, phase-space suppression, and the nuclear form factor.
}
\end{figure}

In Fig.~\ref{fig:prod_xsec_mode}, we show the mode-by-mode production cross section as a function of the KK graviton mode number. The position of the peak reflects the competition between the increasing graviton-photon overlap for modes that probe the intermediate brane and the kinematic and nuclear form-factor suppression that becomes important for heavier states.
This behavior has a simple physical interpretation. For small $n$, the corresponding graviton masses satisfy $m_n^G\ll \Lambda_\gamma$. The KK graviton profiles then remain on their small-$z$ tails throughout the photon interval, so their overlap with the visible photon sector is suppressed. As $n$ increases and $m_n^G$ approaches $\Lambda_\gamma$, the profiles develop appreciable support near the intermediate brane, and the production rate rises. At still larger mode numbers, however, the graviton profiles become increasingly oscillatory over the interval probed by the photon. In addition, phase-space suppression and the nuclear form factor become increasingly important once $m_n^G$ approaches the characteristic incoming photon energy. Consequently, the mode-by-mode production cross section is expected to rise from the lowest modes, reach a maximum for kinematically accessible modes with $m_n^G\sim\Lambda_\gamma$, and then decrease for sufficiently large $n$.

\subsection{Cascade decays}
Once produced, a GeV-scale KK graviton can de-excite through several classes of channels. The produced KK graviton state sits high in a very dense tower, so decays that keep the energy flow within the deep IR are usually favored over direct decays back to visible photons. The main decay channels can be summarized as follows:
\begin{enumerate}[leftmargin=*]
\item \textit{Intratower graviton decays:}
\begin{equation}
G_n \rightarrow G_m G_{\ell}
\end{equation}
with $m+\ell<n$ kinematically and with the exact decay widths determined by the overlap integrals of the three graviton wavefunctions.  Because the graviton profiles are all localized toward the deep IR, these overlap factors are generically large for transitions among nearby KK levels.  For a GeV-scale produced mode one has $n\gg 1$, so there is a large multiplicity of lower-level graviton final states.

\item \textit{Decay to radion final states:}
\begin{equation}
G_n \rightarrow r r, \qquad G_n \rightarrow G_m r 
\end{equation}
whenever $m_n^G>m_m^G+m_r$. 
Here $r$ denotes specifically the lightest radion of the last brane interval, namely the scalar mode associated with fluctuations of the region between $y=y_{\gamma}$ and $y=y_{\mathrm{IR}}$.
The overlap integrals are again localized toward the deep IR region and can provide an efficient route for a heavy KK graviton to de-excite. Because it is localized toward the deepest IR brane, it couples appreciably only to the deep
gravitational sector and has negligible direct coupling to the visible photon and electroweak sectors in the regime of interest. Consequently, decays to radions should be treated phenomenologically as invisible decays: once a heavy KK graviton deposits energy into these radions, that part of the cascade is effectively lost to the detector.

\item \textit{Heavier scalar-KK final states:}
In a stabilized construction, there can also be heavier scalar excitations, denoted collectively by $\phi_p$, associated with the GW stabilization. Representative channels are
\begin{equation}
G_n\to r\,\phi_p,
\qquad
G_n\to G_m\,\phi_p,
\qquad
G_n\to \phi_p\phi_q.
\end{equation}
These are hidden-sector scalar excitations whose couplings are again controlled by overlap integrals in the deep IR.
Once produced, the heavier scalar modes $\phi_p$ are themselves expected to decay further through channels such as
\begin{equation}
\phi_p \to G_m G_\ell,
\qquad
\phi_p \to r\,G_m,
\qquad
\phi_p \to r\,r,
\qquad
\phi_p \to \phi_q G_m,
\end{equation}
whenever allowed by phase space. The exact branching fractions are dictated by the overlap integrals but the qualitative effect is that scalar intermediate states continue to redistribute the energy of the initially produced higher mode KK graviton into progressively lighter hidden-sector modes. Branches containing the light radion-like mode again correspond to invisible energy flow into the deep IR sector in the sequestered benchmark considered here.

\item\textit{Direct visible decay to photons}
\begin{equation}
G_n \rightarrow \gamma \gamma
\end{equation}
Although this coupling is large enough to allow photon-induced production in fixed-target or beam-dump environments, it is typically not the dominant decay mode for the higher excited KK graviton modes.  The photon overlap only probes the interval up to $y=y_{\gamma}$, whereas graviton self-interactions and radion/scalar couplings receive their largest support from the deeper IR region. In addition, the visible diphoton channel is only a single final state, while the hidden-sector cascade has access to a large multiplicity of lower-level graviton and scalar final states. The direct branching ratio to $\gamma\gamma$ is therefore expected to be small for the higher KK graviton modes.
\end{enumerate}

These considerations imply a simple qualitative picture. High-lying gravitons produced in fixed-target or beam-dump environments can cascade through a network of hidden-sector transitions involving lower KK gravitons, the last-interval radion, and, when open, heavier scalar modes. The net effect is a progressive transfer of invariant mass into lighter states of the deep gravitational sector, while the radion-containing branches are effectively invisible to the detector.

\subsection{Terminal diphoton decays}

If the terminal KK graviton mode lies below the two-radion threshold, i.e., $m_1^G < 2m_r$, then the decay $G_1\to rr$ is kinematically forbidden. In the spin-2-dominated benchmark considered here, the leading visible decay of the terminal state is then $G_1\to\gamma\gamma$, which controls its lifetime. The lightest KK graviton can therefore be long-lived and may travel a macroscopic distance before decaying into a pair of photons.

The corresponding partial width is
\begin{equation}
\Gamma\!\left(G_1\to\gamma\gamma\right)
\simeq
\frac{(m_1^G)^3}
{80\pi\,\left(\Lambda_{\gamma\gamma,{\rm eff}}^{(1)}\right)^2}.
\end{equation}
The proper decay length is
\begin{equation}
c\tau_{G_1}
\simeq
\frac{80\pi\,\hbar c}{(m_1^G)^3}
\left(\Lambda_{\gamma\gamma,{\rm eff}}^{(1)}\right)^2 .
\end{equation}
For a boosted terminal graviton, the decay length in the laboratory frame is $\ell_{\rm lab} = \beta\gamma\,c\tau_{G_1}$. 
Depending on the warped scales and the overlap-induced effective coupling, $\ell_{\rm lab}$ can naturally fall in the meter-to-hundreds-of-meters range, making displaced diphoton decays a characteristic signature at beam-dump, fixed-target, and long-baseline near-detector experiments.

\section{Summary and Conclusions}\label{sec:Conclusion}

In this work, we have explored the possibility that warped extra dimensions may give rise to observable signatures at the intensity frontier rather than only through TeV-scale resonances at energy-frontier colliders. We considered an extended warped construction in which the gravitational sector propagates to a deep infrared region with warped scale $\Lambda_{\rm IR}\sim \mathcal{O}({\rm MeV})$. This gives rise to a densely spaced KK graviton tower with MeV-scale mass splittings. To make this light gravitational tower accessible to GeV-scale beam environments, we introduced an intermediate vector brane and studied a benchmark realization in which the hypercharge gauge field propagates down to this intermediate scale, while the Higgs sector remains tied to a higher warped scale.

A key feature of this construction is that the coupling of the KK graviton tower to visible photons is controlled by wave-function overlap in the extra dimension. This differs qualitatively from ordinary light-mediator simplified models, in which the mass and coupling are often treated as independent parameters. In the present framework, the portion of the graviton tower that couples most efficiently to photons is determined geometrically by the endpoint of the vector interval. Modes with masses well below the intermediate scale remain on the small-argument tail of the graviton wave function over the photon interval and couple weakly, whereas modes with masses comparable to the intermediate scale can have enhanced overlap. As a result, GeV-scale photons in fixed-target or beam-dump environments can preferentially produce heavier KK gravitons rather than the lightest modes in the tower.

The resulting phenomenology is characterized by a sequence of processes that is uncommon in standard intensity-frontier light-mediator searches. First, secondary photons produced in a target or beam dump can undergo a Gertsenshtein-like conversion into KK gravitons in the electromagnetic field of the target nuclei. The corresponding mode-by-mode production rate reflects a competition between the increasing photon-graviton overlap for modes probing the intermediate brane and the kinematic and nuclear-form-factor suppression of heavier states. Second, the produced GeV-scale KK gravitons typically sit high in a dense tower and can de-excite through intratower decays into lighter gravitons and, when kinematically allowed, radion or stabilization-sector scalar modes. These hidden-sector cascades redistribute the energy of the initially produced state into progressively lighter degrees of freedom in the deep infrared region. Finally, if the terminal spin-2 state lies below the relevant radion thresholds, its leading visible decay can be $G_1\to\gamma\gamma$. The terminal KK graviton can then be long lived, with a boosted decay length that naturally falls in the macroscopic range for appropriate choices of the warped scales and overlap-induced couplings.

This framework therefore predicts a distinctive class of intensity-frontier signals: preferential production of heavier KK modes, intratower showering in a low-scale gravitational sector, and displaced electromagnetic decays of a terminal state. Such signatures are qualitatively different from those of a single dark photon, ALP, scalar, sterile neutrino, or simplified spin-2 mediator. They motivate dedicated studies of tower-summed production rates, cascade branching patterns, displaced diphoton signatures, and detector-level event topologies at fixed-target, beam-dump, and long-baseline near-detector experiments.

We also examined the main model-building tension associated with this visible hypercharge realization. The same hypercharge field that provides the photon portal to the low-scale gravitational tower also gives rise to a hypercharge KK tower with potentially dangerous couplings to Higgs-brane fields and SM fermions. We showed how BLKTs can suppress these couplings while preserving an appreciable photon-graviton overlap. In particular, the excited hypercharge KK couplings to Higgs-brane matter can be reduced through wave-function distortion and through the normalization of the hypercharge zero mode. The viable parameter space must be chosen with care, so that the photon zero mode retains a positive kinetic term and no spurious or ghostlike vector states are introduced. Within this controlled phenomenological regime, the hypercharge tower can remain tied to the intermediate scale while its direct impact on precision observables is suppressed.

Several directions remain open. A more complete treatment of the radion and stabilization sector would be needed to determine the detailed cascade branching fractions and the conditions under which the terminal spin-2 state dominates the visible phenomenology. A full precision-electroweak analysis would require diagonalizing the neutral vector tower and computing the induced shifts in $Z$-pole and low-energy neutral-current observables. It would also be interesting to develop the more secluded dark-$U(1)_D$ realization, in which a dark vector field rather than hypercharge probes the intermediate brane and communicates with the SM through kinetic mixing. Finally, detailed experimental studies, including realistic photon spectra, detector geometry, backgrounds, and reconstruction efficiencies, are necessary to assess the discovery reach of specific facilities.

The broader lesson is that multi-scale warped geometries can organize light new physics in a way that is not captured by ordinary four-dimensional portal benchmarks. By separating the Higgs, vector, and gravitational endpoints in the extra dimension, one can obtain a low-scale gravitational tower whose masses, couplings, and cascade structure are correlated by geometry. This opens a new route for testing warped extra dimensions at the intensity frontier and provides a benchmark framework for future studies of
geometry-controlled light new physics.

\section*{Acknowledgments}\label{sec:Acknowledgements}
We thank Kaustubh Agashe, Seong Chan Park, Maxim Perelstein, and Lorenzo Ricci for insightful and helpful discussions. The work of DK and AV is supported in part by NSF Grant No. PHY-2609913.
The work of DS is supported by DOE Grant DESC0010813.  DK, DS, and AV gratefully acknowledge the Center for Theoretical Underground Physics and Related Areas (CETUP*), the Institute for Underground Science at Sanford Underground Research Facility (SURF), and the South Dakota Science and Technology Authority for hospitality and financial support, as well as for providing a stimulating environment during the completion of this work. 

\bibliography{ref}

\end{document}